\documentclass{aastex63}
\usepackage{amsmath}
\let\tablenum\relax
\usepackage{siunitx}

\begin{document}

\title{Estimation of Key Sunquake Parameters through Hydrodynamic Modeling and Cross-correlation Analysis}

\correspondingauthor{John T. Stefan}
\email{jts25@njit.edu}

\author{John T. Stefan}
\affiliation{Department of Physics, New Jersey Institute of Technology,
University Heights, Newark, NJ 07102}

\author{Alexander G. Kosovichev}
\affiliation{Department of Physics, New Jersey Institute of Technology,
University Heights, Newark, NJ 07102}
\affiliation{NASA Ames Research Center,
Moffett Field, Mountain View, CA 94035}


\begin{abstract}
    Sunquakes are one of the more distinct secondary phenomena related to solar flares, where energy deposition in the lower layers of the Sun's atmosphere excites acoustic waves easily visible in Helioseismic and Magnetic Imager (HMI) Dopplergrams. We explore two possible excitation mechanisms of sunquakes in the context of the electron beam hypothesis: an instantaneous transfer of momentum and a gradual applied force due to flare eruption. We model the sunquake excitation and compare with five observed sunquake events using a cross-correlation analysis. We find that at least half the events studied are consistent with the electron beam hypothesis and estimate the energy required to excite the sunquakes to be within the range determined by previous studies.
\end{abstract}
\section{Introduction} 

Solar flares represent some of the most energetic phenomena observed in the Solar System. It is well-understood that solar flares result from the reconfiguration of the local magnetic field towards a lower-energy state \citep{Aschwanden_magE,Ulyanov_flare}, which drives particle acceleration and other secondary processes. A comprehensive understanding of energy transport in the solar atmosphere requires close inspection of these secondary processes, which distribute the energy released in flares at coronal heights to regions lower in the atmosphere. Among these processes are sunquakes, which are impulsive seismic events observed in the lower solar atmosphere. While the effects of sunquakes are easily seen in observations - first detected by expanding ripples in MDI dopplergrams \citep{Kosovichev1998,Kosovichev_SQ} - the mechanism by which they are generated has yet to be determined.

Magnetic reconnection in the corona has been identified as the main driver of energy release during solar flares \citep{Anti_reconn,Kopp_reconn}, and electron beams accelerated by this reconnection process are suggested as a possible means of exciting sunquakes \citep[e.g.][]{Sharykin_SQ,Macrae_SQ,Pedram_SQ}. In particular, the thick-target model \citep[e.g.][]{Fisher_TT} suggests that the electron beam strikes a stationary chromosphere, and the resulting thermalization of the electrons can be observed as a sharp increase in hard X-ray (HXR) emission. If the electron beam hypothesis is correct, then there should be some correlation between HXR emission as a function of time and the time of sunquake excitation.

Sunquakes themselves are studied as part of helioseismology, which is concerned with the propagation of acoustic waves of the Sun. The discipline separates the acoustic waves into three separate modes: pressure waves, also called p-modes; gravity and buoyancy waves, also called g-modes; and the f-mode, a surface propagating gravity wave which separates p- and g-modes in the dispersion relation. In power spectra of the Sun's radial velocity field, the p-modes produce a distinct banded structure identified as the resonant modes between 1 and 5 mHz. Non-resonant acoustic waves above 5 mHz are usually referred to as pseudo-modes, as they appear as a similar banded structure in power spectra. Sunquakes generally produce higher frequency waves due to the short duration of excitation, and pseudo-modes are often observed following sunquakes because of this.

Previous studies of sunquakes have identified the photosphere or lower chromosphere as possible locations of sunquake excitation \citep{KosZhar1995,Zharkova,Sharykin_SQ,Riuzhu}. Furthermore, it appears that heating in these regions and the subsequent shocks can provide enough energy to excite the acoustic waves \citep{Kosovichev2015,KosZhar1995}. \citet{Zharkova} find that electron, proton, or mixed-particle beams have the potential to provide the necessary heating, that proton beams may penetrate up to 300 km below the quiet-sun photosphere, and that electron beams can penetrate as far as 5000 km below the photosphere. An alternative mechanism was suggested by \citet{Fisher2012}, who argued that flare eruptions may cause a perturbation of the Lorentz force balance and drive sunquake excitation.

In this work, we will examine the possibility of an electron beam and an applied external force as a sunquake excitation mechanism. We construct a hydrodynamic model and test two types of excitation mechanisms: an instantaneous transfer of momentum to the surrounding atmosphere - analogous to the shock excited by the thermalization of the electron beam - and a more gradual transfer of energy modeled as an applied external force. While the first mechanism is more applicable for strong flares, where the emission is more impulsive, the second type may help to explain sunquakes generated by magnetic field perturbations. We place these sources of excitation in the low chromosphere and photosphere, where it is suspected that sunquakes are excited.

In addition to testing various mechanisms, we also account for changes in the atmosphere through which resulting acoustic waves travel. In higher density regions, such as the chromosphere and photosphere, changes to acoustic wave propagation due to MHD effects are expected to be small as these contributions are relevant only when the Alfv\'en velocity is close to or greater than the acoustic sound speed. This occurs for low densities, but also for very large magnetic fields. The observed sunquakes we use to compare with our model propagate outside their respective active regions, so an MHD treatment of the acoustic waves is not necessary. We do, however, consider the effects of acoustic wave propagation through a damping medium by employing a damping scheme based on wavenumber, which we explore in more depth in section \ref{damping}.

\section{Governing Equations and Methods}
\subsection{Governing Equations}
To construct the model, we begin with the compressible mass continuity and momentum equations
\begin{equation}\label{mass}
\dfrac{D \rho}{D t} + \rho\nabla\cdot\mathbf{v} = 0
\end{equation}
and
\begin{equation}\label{mom}
\rho \dfrac{D \mathbf{v}}{D t} = -\mathbf{\nabla}P + \rho\mathbf{g} + \mathbf{F}_{ext},
\end{equation}
where D/Dt is the material derivative and is equivalent to
\[\dfrac{D}{Dt}=\dfrac{\partial}{\partial t}+\mathbf{v}\cdot\mathbf{\nabla},\]
which includes changes in time and also gradients advected by the velocity field. We assume that wave fronts travel through a given point in the system quickly enough such that there is no exchange of heat between the point and the surroundings, and the entropy of the system should therefore be unaffected, i.e. propagation is adiabatic with the condition

\begin{equation}\label{adi}
\dfrac{DS}{Dt}=\dfrac{D}{Dt}\left(\dfrac{P}{\rho^{\gamma}}\right)=0.
\end{equation}

We will consider linear Eulerian perturbations to the above equations, such that the background state is assumed to be time-independent and contains only radial variations: $\rho\rightarrow \rho_{0}(r) + \rho^{\prime}(r,\vartheta,\varphi,t)$. We choose the case where there are no background velocity fields, and so the fluid velocity is itself a perturbation: $\mathbf{v}\rightarrow\mathbf{v}(r,\vartheta,\varphi,t)$. Furthermore, we separate the velocity into radial and horizontal components, where the horizontal component contains both the $\hat{\theta}$ and $\hat{\phi}$ components. Finally, we assume that the radial and $\vartheta$,$\varphi$ dependence can be separated as
\[
a(r,\vartheta,\varphi)=\sum_{l,m}a_{l,m}(r)Y_{l}^{m}(\vartheta,\varphi),
\]
where the $Y_{l}^{m}$ are the spherical harmonics.

Beginning with the equation for mass conservation, equation \ref{mass}, we expand the variables in terms of their background and perturbed quantities and keep only linear terms
\[
\dfrac{\partial \rho^{\prime}}{\partial t} + \mathbf{v}\cdot\mathbf{\nabla}\rho_{0}+\rho_{0}\mathbf{\nabla}\cdot\mathbf{v} =0.
\]
We introduce the horizontal velocity $\mathbf{v_{h}}$, the velocity vector in the $\hat{\theta}$ and $\hat{\phi}$ directions, which describes flow along a spherical shell. Expanding derivatives and separating the radial and horizontal divergence ($\mathbf{\nabla_{h}}\cdot$), we arrive at
\[
\dfrac{\partial \rho^{\prime}}{\partial t}+v_{r}\dfrac{\partial\rho_{0}}{\partial r}+\rho_{0}\left(\dfrac{1}{r^{2}}\dfrac{\partial}{\partial r}\left(r^{2}v_{r}\right)+\dfrac{1}{r}\nabla_{h}\cdot\mathbf{v_{h}}\right)=0.
\]
We choose the case where the angular dependence of the horizontal component of velocity can be separated as $\sum_{l,m}v_{h,l}(r,t)\mathbf{\nabla_{h}}Y_{l}^{m}(\vartheta,\varphi)$, where $\mathbf{\nabla_{h}}$ is the horizontal gradient which contains the $\theta$ and $\phi$ components of the full gradient operator and $v_{h,l}$ is the scalar horizontal velocity. Expressing the equation with any associated dependences (and summation over $l$ and $m$ is considered implicit) gives
\[
\left[\dfrac{\partial \rho^{\prime}(r,t)}{\partial t} +v_{r,l}(r,t)\dfrac{\partial \rho_{0}(r)}{\partial r}+\rho_{0}(r)\dfrac{1}{r^{2}}\dfrac{\partial}{\partial r}\left(r^{2}v_{r,l}(r,t)\right) \right]\cdot Y_{l}^{m}(\vartheta,\varphi)
+\rho_{0}(r)\left[\dfrac{v_{h,l}(r,t)}{r}\right]\nabla_{h}^{2}Y_{l}^{m}(\vartheta,\varphi)=0.
\]

It is convenient to choose the case where the sunquake is centered over the pole at $\vartheta=0$, as the system is then azimuthally symmetric and the spherical harmonics reduce to the associated Legendre polynomials $P_{l}(\cos\vartheta)$. In this case, the term $\nabla_{h}^{2}P_{l}$ is equivalent to $-L^{2}P_{l}=-l(l+1)P_{l}$ for any positive integer $l$ and no angular derivatives need to be explicitly computed. We use the shorthand $x_{l}=A_{l}x$, where the variable $x_{l}$ is any variable $x$ with its associated Legendre coefficients $A_{l}$. Making the appropriate substitutions and dividing through by $\rho_{0}$ we arrive at
\begin{equation}\label{lin_mass}
\dfrac{\partial \bar{\rho}_{l}}{\partial t}+v_{r,l}\dfrac{\partial \ln\rho_{0}}{\partial r}+\dfrac{1}{r^{2}}\dfrac{\partial}{\partial r}\left(r^{2}v_{r,l}\right)-\dfrac{L^{2}}{r}v_{h,l}=0,
\end{equation}
where $\bar{\rho}=\rho^{\prime}/\rho_{0}$ is the normalized perturbation to density.

We move now to the energy equation (Equation \ref{adi}); expanding the material derivative and expressing the variables in terms of background and perturbed quantities yields
\[
\dfrac{1}{\gamma P_{0}}\left[\dfrac{\partial P^{\prime}}{\partial t}+\mathbf{v}\cdot\mathbf{\nabla}P_{0}+\mathbf{v}\cdot\mathbf{\nabla}P^{\prime}\right] = \dfrac{1}{\rho_{0}}\left[\dfrac{\partial \rho^\prime}{\partial t}+\mathbf{v}\cdot\mathbf{\nabla}\rho_{0}+\mathbf{v}\cdot\mathbf{\nabla}\rho^{\prime}\right],
\]
where we have made the approximations $1/\gamma(P_{0}+P^\prime)\approx1/\gamma P_{0}$ and $1/(\rho_{0}+\rho^{\prime})\approx1/\rho_{0}$ since $P^{\prime}\ll P_{0}$ and $\rho^{\prime}\ll\rho_{0}$. We now remove non-linear terms, keeping in mind that background pressure and density vary only in the radial direction
\[
\dfrac{1}{\gamma P_{0}}\dfrac{\partial P^{\prime}}{\partial t}+v_{r}\left(\dfrac{1}{\gamma P_{0}}\dfrac{\partial P_{0}}{\partial r}-\dfrac{1}{\rho_{0}}\dfrac{\partial \rho_{0}}{\partial r}\right)=\dfrac{1}{\rho_{0}}\dfrac{\partial \rho^{\prime}}{\partial t}.
\]

\begin{figure}[!hbt]
\centering
\includegraphics[width=0.5\linewidth]{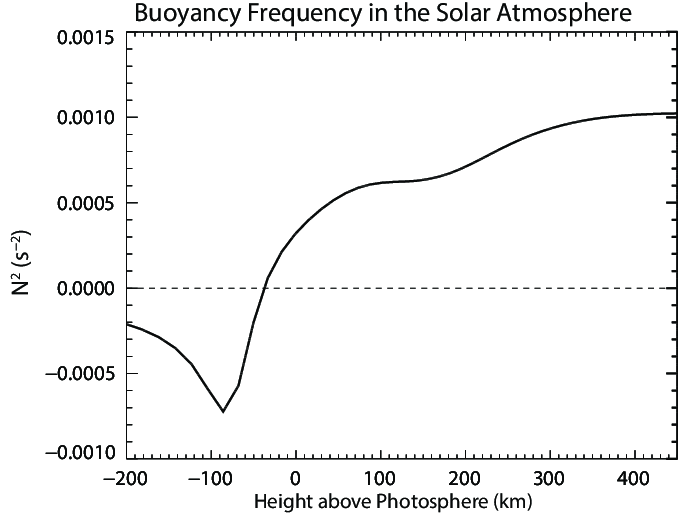}
\caption{Square of the buoyancy frequency as a function of height in the solar atmosphere, with z=0 km being the base of the photosphere. Negative values correspond to unstable g-mode propagation and positive values indicate stable regions of g-mode propagation.}
\label{boyfreq}
\end{figure}

Note that the Brunt-V\"ais\"al\"a frequency (N), or buoyancy frequency, is defined as
\[
N^{2}=g\left(\dfrac{1}{\gamma P_{0}}\dfrac{\partial P_{0}}{\partial r}-\dfrac{1}{\rho_{0}}\dfrac{\partial \rho_{0}}{\partial r}\right).
\]
The buoyancy frequency determines the frequency and stability of g-modes; in the solar core and atmosphere, this frequency is real-valued (as in Figure \ref{boyfreq}) and produces stable g-modes. In the convective region, this frequency is imaginary, and corresponds to convective instabilities; to ensure the stability of wave modeling, we set $N^{2}$ to be zero in this region. We make the buoyancy frequency substitution in the energy equation
\[
\dfrac{1}{\gamma}\dfrac{\partial\bar{P}}{\partial t}+v_{r}\dfrac{N^{2}}{g} = \dfrac{\partial \bar{\rho}}{\partial t},
\]
where $\bar{P}=P^{\prime}/P_{0}$ is the normalized perturbation to pressure. Replacing the normalized density time-derivative with Equation \ref{lin_mass}, and substituting the Legendre polynomials, we find the final form for the energy equation
\begin{equation}\label{lin_energy}
\dfrac{\partial \bar{P}_{l}}{\partial t}  +\gamma\left[v_{r,l}\left(\dfrac{N^{2}}{g} + \dfrac{\partial \ln\rho_{0}}{\partial r}\right) +\dfrac{1}{r^{2}}\dfrac{\partial }{\partial r}\left(r^{2}v_{r,l}\right)- \dfrac{L^{2}}{r}v_{h,l}\right]=0 .
\end{equation}

Lastly, we begin our treatment of the momentum equation by first separating the radial and horizontal components, and performing the same expansion as for Equations \ref{lin_mass} and \ref{lin_energy}. Looking first at the radial momentum equation
\[
\rho_{0}\dfrac{\partial v_{r}}{\partial t}=-\dfrac{\partial P_{0}}{\partial r}-\dfrac{\partial P^{\prime}}{\partial r}+\rho_{0}g+\rho^{\prime}g+F_{ext,r}.
\]

We assume that the background state is in hydrostatic equilibrium, which satisfies the relation $\partial P_{0}/\partial r + \rho_{0}g = 0$, and these terms then fall out of the equation. Dividing through by $\rho_{0}$ to isolate the time-derivative and substituting the Legendre polynomials leaves us with
\begin{equation}\label{lin_radmom}
\dfrac{\partial v_{r,l}}{\partial t}=-\dfrac{1}{\rho_{0}}\dfrac{\partial P^{\prime}_{l}}{\partial r}+\bar{\rho}_{l}g+\dfrac{F_{ext,r,l}}{\rho_{0}}.
\end{equation}

The horizontal component of the momentum equation has the original form
\[
\rho_{0}\dfrac{\partial v_{h}}{\partial t}=\dfrac{1}{r}\nabla_{h}P^{\prime}+F_{ext,h}.
\]
When expanding the variables in terms of radial and angular dependence, we note that $P^{\prime}$ varies only in time and in radius, so the horizontal gradient is moved to the right and applied to the Legendre polynomials. We assume that the horizontal component of external forces can be decomposed as $F_{ext,h}=\sum_{l}F_{ext,h,l}(r,t)\nabla_{h}P_{l}(\vartheta)$. Since $\nabla_{h}P_{l}$ now appears in every term, it can be neglected during computation but still be used to reconstruct $\vartheta$ dependence. Dividing through by $\rho_{0}$ yields the final form
\begin{equation}\label{lin_hormom}
\dfrac{\partial v_{h,l}}{\partial t}=\dfrac{P^{\prime}_{l}}{r\rho_{0}}+\dfrac{F_{ext,h,l}}{\rho_{0}}.
\end{equation}

The system of equations used to model sunquake propagation is then
\[
\left\lbrace
\begin{array}{l}
\dfrac{\partial \bar{\rho}_{l}}{\partial t}+v_{r,l}\dfrac{\partial \ln\rho_{0}}{\partial r}+\dfrac{1}{r^{2}}\dfrac{\partial}{\partial r}\left(r^{2}v_{r,l}\right)-\dfrac{L^{2}}{r}v_{h,l}=0\\
\dfrac{\partial \bar{P}_{l}}{\partial t}  +\gamma\left[v_{r,l}\left(\dfrac{N^{2}}{g} + \dfrac{\partial \ln\rho_{0}}{\partial r}\right) +\dfrac{1}{r^{2}}\dfrac{\partial }{\partial r}\left(r^{2}v_{r,l}\right)- \dfrac{L^{2}}{r}v_{h,l}\right]=0 \\
\dfrac{\partial v_{r,l}}{\partial t}=-\dfrac{1}{\rho_{0}}\dfrac{\partial P^{\prime}_{l}}{\partial r}+\bar{\rho}_{l}g+\dfrac{F_{ext,r,l}}{\rho_{0}} \\
\dfrac{\partial v_{h,l}}{\partial t}=\dfrac{P^{\prime}_{l}}{r\rho_{0}}+\dfrac{F_{ext,h,l}}{\rho_{0}}
\end{array}\right. .
\]

\subsection{Boundary Conditions}

Acoustic waves which propagate at or above the acoustic cut-off frequency, $\omega_c$, travel through the solar surface and are mostly damped in the atmosphere \citep{Dalsgaard-hseis}. Waves which propagate below this frequency are reflected at the surface and are resonant. To implement this selective resonant behavior in the model, we must choose our boundary conditions carefully. Using Dirichlet boundary conditions leads to reflections at the upper computational domain, and it is not entirely clear how Neumann boundary conditions can be formulated a priori to minimize reflections.

We therefore derive non-reflecting boundary conditions by performing an eigendecomposition of the matrix $\mathit{B}$ which contains the radial derivatives of our system. To simplify calculations, we include only Equations \ref{lin_energy} and \ref{lin_radmom}; this is reasonably justified since $\rho^\prime$ and $P^{\prime}$ are essentially equivalent through Equation \ref{adi}, and Equation \ref{lin_hormom} contains no radial derivatives. The matrix form of the simplified system is given by
\[
\mathit{A}\dfrac{\partial \mathbf{X}}{\partial t}+\mathit{B}\dfrac{\partial \mathbf{X}}{\partial r}=\mathbf{C},
\]
where $\mathit{A}$ is the matrix containing time-derivative coefficients, $\mathbf{X}=[P^{\prime}_{l},v_{r,l}]^{T}$ are the variables, and $\mathbf{C}$ is the vector containing constants. We choose to use the form of Equation \ref{lin_energy} where pressure perturbations are not normalized, and the matrix $\mathit{A}$ is then simply the identity matrix. The radial-derivative matrix $\mathit{B}$ then has the form
\[
\mathit{B}=\begin{bmatrix}
0 & \gamma P_{0} \\
\dfrac{1}{\rho_{0}} & 0
\end{bmatrix},
\]
which has eigenvalues $\lambda_{+}=+c_{s}$ and $\lambda_{-}=-c_{s}$, where $c_{s}=\sqrt{\gamma P_{0}/\rho_{0}}$ is the adiabatic sound speed. With these eigenvalues we obtain the eigenvalue decomposition
\[
\mathit{B}=
\dfrac{1}{2}\begin{bmatrix}
\rho_{0}c_{s} & -1 \\
1 & \dfrac{c_{s}}{\gamma P_{0}}
\end{bmatrix}\begin{bmatrix}
+c_{s} & 0 \\
0 & -c_{s}
\end{bmatrix}
\begin{bmatrix}
\dfrac{c_{s}}{\gamma P_{0}} & 1 \\
-1 & \rho_{0}c_{s}
\end{bmatrix}.
\]
Performing the matrix multiplication within $\mathit{B}$ and keeping terms separate allows us to determine which derivatives correspond to the inward propagating waves. For these terms, we enforce hydrostatic equilibrium for the perturbed quantities ([$\partial P^{\prime}_{l}/\partial r]=-\rho^{\prime} g$ and $[\partial v_{r,l}/\partial r]=0$) so that these derivatives need not be evaluated. The boundary conditions for $R=R_{max}$ used are then
\[
\left\lbrace
\begin{array}{l}

\dfrac{\partial \bar{\rho}_{l}}{\partial t}+v_{r,l}\left(\dfrac{\partial \ln{\rho_{0}}}{\partial r}+\dfrac{2}{r}\right)-\dfrac{L^{2}}{r}v_{h,l}+\dfrac{1}{2}\left(c_{s}\dfrac{\partial \bar{\rho}_{l}}{\partial r}+\dfrac{\partial v_{r,l}}{\partial r}+\dfrac{\bar{\rho}_{l}g}{c_{s}}\right)=0 \\

\dfrac{\partial \bar{P^{\prime}}_{l}}{\partial t} + \gamma\left(v_{r,l}\dfrac{\partial \ln{\rho_{0}}}{\partial r}+v_{r,l}\dfrac{N^{2}}{g}-\dfrac{L^{2}}{r}v_{h,l}\right) +\dfrac{1}{2}\left(c_{s}\dfrac{\partial \bar{\rho}_{l}}{\partial r}+\dfrac{\partial v_{r,l}}{\partial r}+\dfrac{\bar{\rho}_{l}g}{c_{s}}\right)=0 \\

\dfrac{\partial v_{r,l}}{\partial t} + \bar{\rho}_{l}g+\dfrac{1}{2}\left(c_{s}^{2}\dfrac{\partial \bar{\rho}_{l}}{\partial r}+c_{s}\dfrac{\partial v_{r,l}}{\partial r} -\bar{\rho}_{l}g\right)=0 \\

\dfrac{\partial v_{h,l}}{\partial t} + \dfrac{P^{\prime}_{l}}{r\rho_{0}} = 0

\end{array},
\right. 
\]
where we have made the substitution $P^{\prime}_{l}=c_{s}^{2}\rho^{\prime}_{l}$.

\subsection{Numerical Methods}

The governing equations are solved along a radial mesh containing values for radius, background density, background pressure, adiabatic exponent, and gravitational acceleration. The background model used for the simulations is the Standard Solar Model as described by \citet{ssmodel}, which is computed to $R=696.841$ Mm with $501$ grid points.

The time derivatives are approximated by a first order forward difference
\[
\dfrac{\partial y}{\partial t}\approx \dfrac{y^{n+1}_{i}-y^{n}_{i}}{\tau},
\]
where $\tau$ is the time step chosen to satisfy the CFL condition \citep{CFL} $\tau_{CFL}\leq\min[\Delta x/c_{s}]$, corresponding to the travel time of an acoustic wave between the shortest grid point separation $\Delta x$. We choose to use $\tau=0.6\tau_{CFL}$ for stability. A detailed analysis of the stability of the system can be found in the appendix, Section \ref{stab}. The radial derivatives are approximated by a fourth order central difference, which has the following form for a uniform grid
\[
\dfrac{\partial y}{\partial r}\approx \dfrac{y^{n}_{i-2}-8y^{n}_{i-1}+8y^{n}_{i+1}-y^{n}_{i+2}}{12\Delta x}.
\]

We use a non-uniform grid, however, and the appropriate coefficients and grid separations are substituted. We employ a staggered mesh scheme, where the pressure, density, and horizontal velocity variables are placed on "body" points, and the radial velocity variable is placed on "edge" points halfway between two body points; i.e. $r_{b,i} < r_{e,i} < r_{b,i+1}$. Values for pressure and density are computed first, followed by the radial and horizontal velocities using the updated pressure and density values. The velocity radial derivative at the upper boundary is evaluated using a second order central difference, as the boundary body point has edge points both above and below. On uniform grid, this approximation has the form
\[
\dfrac{\partial y}{\partial r} \approx \dfrac{y_{i+1}^{n}-y_{i-1}^{n}}{2\Delta x}
\]
and the appropriate substitutions are made for implementation on a staggered mesh. The perturbed density radial derivative at the upper boundary is evaluated using the same difference, though a ghost point is implemented beyond the boundary so that a central difference can be used. The pressure and density variables at the ghost point are evolved in time using the boundary conditions, and radial derivatives are approximated with a second order one-sided difference, of the form
\[
\dfrac{\partial y}{\partial r} \approx \dfrac{3y_{i}^{n}-4y_{i-1}^{n}+y_{i-2}^{n}}{2\Delta x}.
\]

The discretized governing equations and time advancement scheme are written in Fortran, and parallelized using the MPI (message-passing interface) library. The code is run on 504 nodes of the Pleiades supercomputer at NASA Ames Research Center; we compute the solutions to the governing equations up to angular degree $l=6000$, so each node runs approximately twelve iterations of the program in series. After the computation is completed, the data is stored as function of radius, time, and angular degree $l$. Damping is applied to each angular order (as described in the following section) and a spherical harmonic transformation takes the data from a function of $l$ to a function of angle $\theta$.

\subsection{Damping by Wavenumber}\label{damping}

Our governing equations have so far neglected any effects from viscous damping, as the plasma viscosity is highly dependent on ionization, temperature, and magnetic field strength \citep{visc}, which are outside the scope of this research. We instead choose to use an ad hoc damping scheme, where acoustic waves are damped by wavenumber, and appropriate parameters are derived from observation. We assume that acoustic waves traveling with frequency $\omega_{0}$ have time-dependence of the form $\Psi(t)=A\exp[i\omega_{0}t]\exp[-\alpha_{l} t]$, where $\alpha_{l}$ is a damping parameter dependent on the angular degree $l$ associated with the frequency $\omega_{0}$. The angular degree and horizontal wavenumber are closely related by $k_{h}=\sqrt{l(l+1)}/R_{\odot}$, and we choose to evaluate $\alpha_{l}$ in terms of the angular degree since our governing equations are solved in this way as well.

The power spectrum of the signal is dependent on frequency as
\[
P=\dfrac{A}{4\pi^{2}}\dfrac{1}{(\omega_{0}-\omega)^{2}+\alpha_{l}^{2}}.
\]
Note that the power is maximized at the frequency $\omega_{0}$. Evaluating the power at the upper frequency ($\omega_{+}$) of the full-width, half-maximum ($FWHM=2(\omega_{0}-\omega_{+}$) and equating with $P_{max}$ shows the damping parameter is $\alpha_{l}=(1/2)(FWHM)$. The associated damping time ($\tau_{l}$) is the inverse of $\alpha_{l}$. For simplicity, we assume that the damping time varies with $l$ as
\[
\tau_{l}=\tau^{\star}\left(\dfrac{l}{l^{\star}}\right)^{\gamma},
\]
where $l^{\star}$ is an arbitrary base angular degree, $\tau^{\star}$ is the damping time of the base angular degree, and $\gamma$ is a power-law exponent derived from observations.

\begin{figure}[htb]
\includegraphics[width=\textwidth]{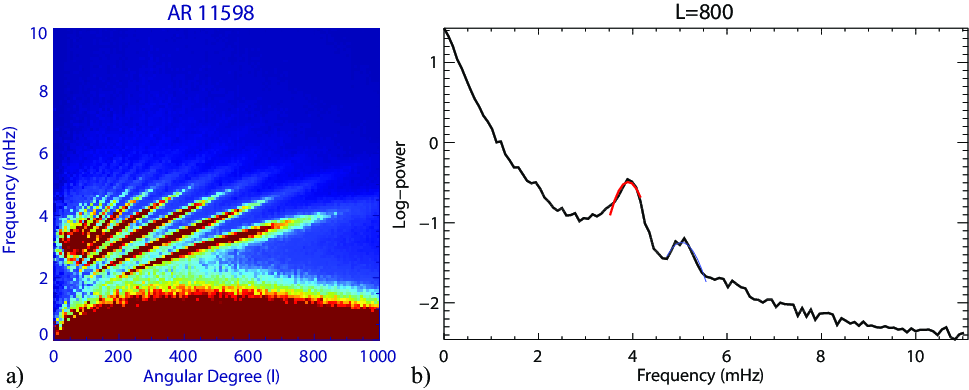}
\caption{(a) Power spectrum obtained from the Dopplergram series of AR 11598. Redder colors indicate greater power, bluer colors indicate lesser power}. (b) Power spectrum of angular degree $l=800$. The first peak (in red) corresponds to $\tau_{800}=408$ seconds, and the second peak (in blue) corresponds to $\tau_{800}=422$ seconds.\label{powerspec}
\end{figure}

We consider three damping cases: quiet-sun damping, active-region damping, and no damping. For the first case, we use p-mode data from \citet{pmodefreq}, in which a three day full-disk dopplergram series is used to compute an azimuthally averaged power spectrum. The damping times are derived from the FWHM data of the set, and the obtained damping times are fit to a power law using a least-squares algorithm to derive the index. We find $\gamma=-0.723$ and for the chosen base $l=1000$, the damping time is 843 seconds. For active-region damping, we assume that only the damping times change, and the exponent $\gamma$ remains constant. The power spectrum of a three-hour dopplergram series of AR 11598 (Figure \ref{powerspec}a) is used to obtain the damping time for the base angular order $l=800$, which is chosen instead of $l=1000$ as this degree is not well-resolved from the background. There are two discernible peaks in the $l=800$ power spectrum (Figure \ref{powerspec}b), and fitting with a Gaussian profile yields a damping time of 408 seconds for the first peak and 422 seconds for the second; we use the average of 415 seconds as the base $\tau^{\star}$. Since the solution is stored as function of angular degree $l$, we use these parameters to apply the damping to the respective angular order. Once the damping has been applied, a spherical harmonic transformation is used to express the solution as function of the angle $\theta$ instead of the angular degree $l$.

\subsection{Form of Excitation Mechanisms}

We consider two types of excitations, initial conditions for momentum impact and applied external forces. The momentum initial conditions are time-independent, and the external force excitations are time-dependent, which we use to simulate gradual processes. Both the momentum and force are directed in the downward radial direction, and have Gaussian radial and angular dependences; the time dependence of the force excitations are also Gaussian. In the general case, the force impact is modeled with the form
\begin{equation*}
F_{r}(r,\vartheta,t) =A  \exp\left[-\dfrac{(r-r_{0})^{2}}{2\sigma_{r}^{2}}\right] \exp\left[-\dfrac{\vartheta^{2}}{2\sigma_{\vartheta}^{2}}\right]\cdot\exp\left[-\dfrac{(t-t_{0})^{2}}{2\sigma_{t}^{2}}\right],
\end{equation*}
where $\sigma_{r}$, $\sigma_{\vartheta}$, and $\sigma_{t}$ are parameters related to the full-width half-maximum of the respective Gaussians, $r_{0}$ is the radial center of the excitation source, $t_{0}=T/2$ is the central time chosen to be half the duration of the textbf{excitation}, the duration T = 200 s, $\sigma_{r} = 0.2$ Mm, $\sigma_{t}=T/4=50$ seconds, and the amplitude $A$ is arbitrary and negative, chosen to be $-0.001$ dyn cm$^{-3}$. Momentum initial conditions are applied to the radial velocity equation with the form
\begin{equation*}
    V_{r}\left(r,\vartheta,0\right)=\dfrac{A}{\rho_{0}\left(r\right)}  \exp\left[-\dfrac{(r-r_{0})^{2}}{2\sigma_{r}^{2}}\right] \exp\left[-\dfrac{\vartheta^{2}}{2\sigma_{\vartheta}^{2}}\right];
\end{equation*}
again, the amplitude is arbitrary and negative, though a value of $-100\rho_{0}(r)$ g m s$^{-1}$ is chosen, where $\rho_{0}(r)$ is the background density at a given grid point. Additionally, the $\sigma_{\vartheta}$ parameter is fixed at \num{8.6e-4} radians, which corresponds to a circular area on the surface ($R=695.9906$ Mm) with radius $0.6$ Mm. 

\section{Results}

\subsection{Differences Between Force and Momentum Mechanisms}

\begin{figure}[!bth]
\centering
\includegraphics[width=\linewidth]{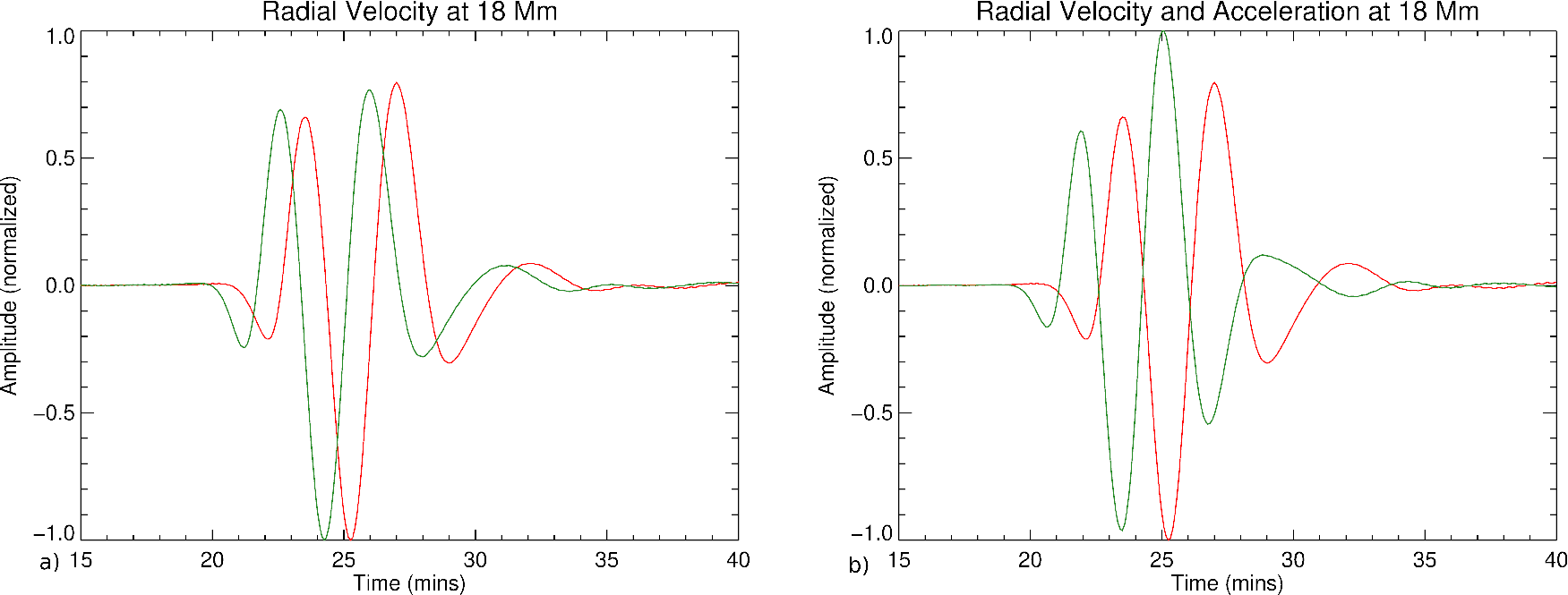}
\caption{(a) Normalized radial velocity measured at a distance of 18 Mm from the excitation source, for momentum (green) and force (red) sources. (b) Normalized radial velocity time-derivative (green) of the momentum excitation and radial velocity of the force mechanism (red).}\label{phasediff}
\end{figure}

We first examine the relationship between force and momentum mechanisms, as we expect there to be a 90 degree phase difference between the time-distance diagrams of the two mechanisms. This phase difference is motivated by the relationship between force and velocity; functionally speaking, force is the time derivative of velocity (multiplied by mass, of course) and ought to be orthogonal to the velocity itself. The force impact has a duration of $T=200$ seconds, and the momentum initial conditions are time-independent. We observe the expected phase difference in time dependence of the radial velocity signal at a distance of 18 Mm from the excitation source, which can be seen clearly in Figure \ref{phasediff}, and note that it persists regardless of excitation depth. Interestingly, the phase difference is actually -90 degrees, as differentiation with respect to time of the signal from the momentum impact yields a 180 degree phase difference (Figure \ref{phasediff}). The signals of the two mechanisms are separated by several minutes, well within HMI resolution, and if the start time of the sunquake is known, then matching the ray path with the sunquake's time-distance diagram may be useful in determining what type of mechanism initiated the quake.

\begin{figure*}[!htb]
\centering
\includegraphics[width=0.9\linewidth]{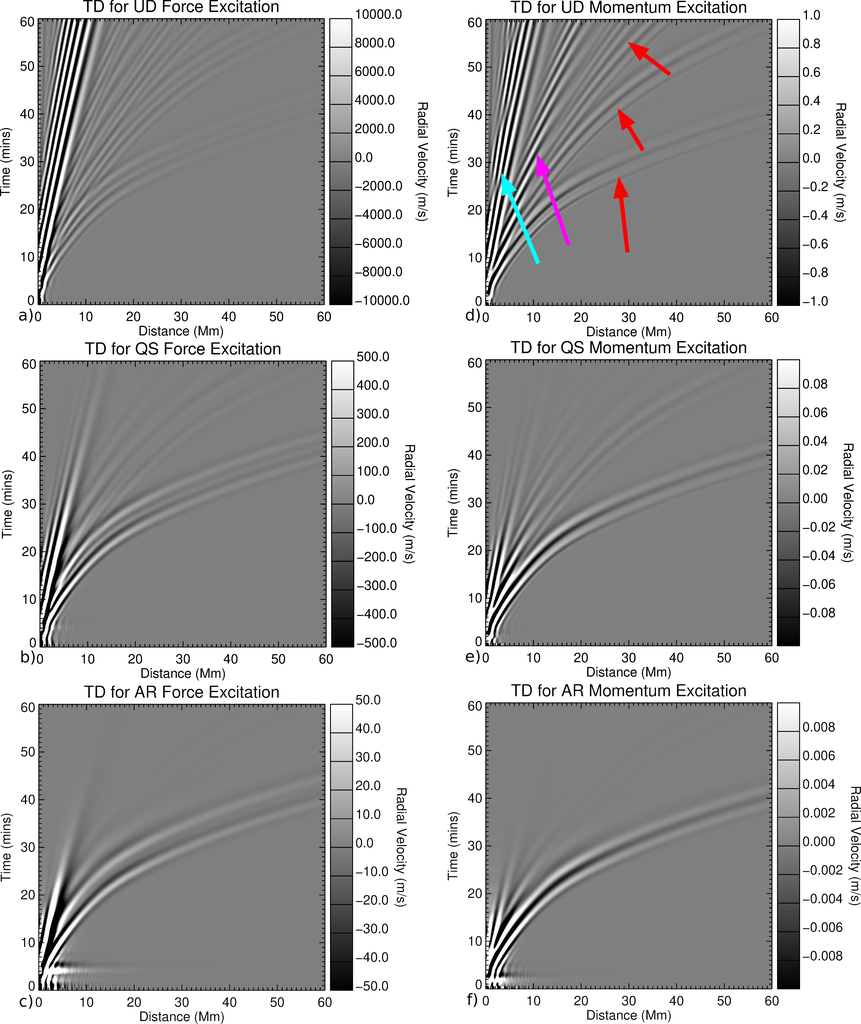}
\caption{(Left column) Time-distance diagrams of radial velocity on the solar surface (R=696 Mm) for force excitations with increasing damping: a) undamped, b) quiet-sun damped, and c) active region damped.  (Right column) Time-distance diagrams of radial velocity on the solar surface (R=696 Mm) for momentum excitations with increasing damping: d) undamped excitation with p-modes (red arrows), f-mode (magenta arrow), and atmospheric acoustic-gravity waves (cyan arrow) highlighted; e) quiet-sun damped; f) active region damped. Darker pixels correspond to more negative velocities, lighter pixels correspond to more positive velocities.}\label{VFcomp}
\end{figure*}

Additionally, we find that the appearance of the time-distance diagrams of the two mechanisms, which display line-of-sight (mostly radial) velocity, are completely distinct (Figure \ref{VFcomp}). For comparison, we center force and momentum sources at R = 696.119 Mm with $\sigma_{r}$ = 0.2 Mm. The first bounce signal of the undamped time-dependent force excitation appears to be longer than that of the momentum excitation, and also has more complicated structure.

We are able to identify the three types of wave propagation: p-modes, the f-mode, and atmospheric acoustic-gravity waves in these time-distance diagrams. The p-modes (red arrows in Figure \ref{VFcomp}d) form the main wavepacket of the sunquake, which travel downwards and are reflected upwards by the sound speed gradient. The f-mode (magenta arrow in Figure \ref{VFcomp}d) is formed by surface gravity waves, and has a distinctive ridge pattern where the phase speed is twice that of the group velocity \citep{Gizon}, and acoustic gravity waves (cyan arrow in Figure \ref{VFcomp}d) are recognized by constant phase and group velocity. Both models produce atmospheric gravity waves, which propagate in a similar fashion to atmospheric gravity waves on Earth (e.g. \citet{row_atmo}). In linear acoustics the solar atmosphere is stable to gravity wave propagation (Figure \ref{boyfreq}), though these waves - especially of such strength - have not been observed. This may be due to break-up caused by convective up-flows and down-flows, which occupy the same frequencies as gravity waves in solar power spectra. The gravity waves in the force excitation model are stronger and more coherent than in the momentum excitation model. The model also reproduces the surface gravity wave (f-mode), which is regularly observed \citep{fmode}, and in the model precedes the shallow water waves. The f-mode is more prominent in momentum excitations and there is longer time delay between f-mode and gravity wave arrival as compared to the force excitations.

As damping is increased, the atmospheric acoustic-gravity waves are affected more heavily than the seismic wave, though the consecutive bounces of the seismic wave and the f-mode show decreases in amplitude. The general phase relationship between momentum-excited seismic signals is preserved, though it appears to be smoothed, with shorter wavelength features blending into the the longer wavelength features. The force excitation model shows similar smoothing and the first bounce signal is stretched into two main packets. This separation into two packets intensifies in the active region damped case, and the acoustic-gravity wave in this case is almost entirely damped. The acoustic-gravity wave damping is also present in the active region damped momentum excitation, though these waves are damped stronger in this case. In both models the f-mode is also entirely damped, and consecutive bounces are barely measurable. We note that there is some corruption of signal near the origin, which is caused by the damping on higher wavenumbers, leading to the loss of small-scale resolution and a blurring effect.

\subsection{Observational Comparisons}

We compare our results with a number of observed sunquakes, which are the events associated with the X1.8 flare in October 2012, the X9.3 flare in September 2017, the X3.3 flare in November 2013, the X1.0 flare in March 2014, and the M1.1 flare in September 2015 (Table \ref{SQtable}) from the sunquake catalog of \citet{Sharykin2019}.

\begin{table}[!htb]
    \caption{Sunquake Events and Relevant Times}\label{SQtable}
    \begin{center}
    \begin{tabular}{|c|c|c|c|c|c|c|c|}
         \hline
         GOES Flare Class & Date & T$_{start}$ (UT) & Lon.$^\dagger$ (deg) & Lat.$^\dagger$ (deg) & T$_{G1}$ (s) & T$_{G3}$ (s) & T$_{BP}$ (s) \\
         \hline \hline
         X1.8 & 2012 Oct 23 & 03:16:30 & 110.3 & -12.7 & +85 & +141 & +45 \\ \hline
         X9.3 & 2017 Sep 6 & 11:57:00 & 122.6 & -9.1 & -82 & -49 & -60 \\ \hline
         X3.3 & 2013 Nov 5 & 22:10:19 & 175.5 & -12.6 & +34 & +76 & +56 \\ \hline
         X1.0 & 2014 Mar 29 & 17:45:00 & 132.5 & +32.0 & +17 & +48 & +135 \\ \hline
         M1.1 & 2015 Sep 30 & 13:15:00 & 108.0 & -21.0 & +86 & +48 & N/A \\ \hline
    \end{tabular}\\
    \hfill \\
    Sunquake events used for comparison, where the start time is the beginning time of the Dopplergram series. The flare relevant times are listed as a time shift relative to the start time of the Dopplergram series. $^\dagger$ Latitude and longitude are given in Carrington heliographic coordinates.
    \end{center}
\end{table}

Time-distance diagrams are produced for these events using data from the Helioseismic and Magnetic Imager (HMI) from Solar Dynamics Observatory (SDO) \citep{Scherrer2012}, after a frequency filter with a Gaussian cut-off is applied to the dopplergram series. The central frequency and width of the filter vary, and are chosen to increase the contrast of the sunquake signal from the background convective noise. Time-distance diagrams are also produced from the model runs, and are treated with an identical filter for comparison. The model runs used in the comparison are separated into two sets, one of momentum excitations and one of force excitations. Each set contains 46 modeled sunquakes with fixed radial width centered along the grid points of the Standard Solar Model, corresponding to the range of R=695.788 Mm to R=696.422 Mm.

In increments of 1 Mm along the horizontal axis, the cross-correlation $\Xi\left(\tau,\xi,d,z\right)$ of the observed sunquake to the set of modeled sunquakes is computed as a function of time-shift ($\tau$) and distance-shift ($\xi$). The signal from the observed sunquake $S$ has dependence on time $t$ and horizontal distance from the excitation source $d$, and the collection of modeled sunquake signals $S^{\prime}$ also has these dependencies with its own distance from the excitation source $d^{\prime}$, and an additional dependence on excitation source depth $z$. The cross-correlation is then dependent on four parameters: time-shift, distance-shift, excitation source depth ($z$), and distance from excitation source ($d$). We initially compute the cross-correlation over time-shift and distance-shift, for each excitation depth and distance from excitation source, as
\[
\Xi\left(\tau,\xi,d,z\right)=\sum_{t}\sum_{d^{\prime}}S\left(t,d\right)S^{\prime}\left(t+\tau,d^{\prime}+\xi,z\right) .
\]
Since a perfect match of observed to modeled sunquake signals will not vary with respect to distance from the excitation source, the cross-correlation is averaged over $d$
\[
\bar{\Xi}\left(\tau,\xi,z\right)=\dfrac{1}{N_{d}}\sum_{d}\Xi\left(\tau,\xi,d,z\right),
\]
where $N_{d}$ is the number of pixels along the horizontal axis. Finally, the set of cross-correlations which are maximized with respect to distance-shift are identified, and the resulting cross-correlation function $\Xi^{\star}\left(\tau,z\right)$ is dependent only on time-shift and excitation depth
\[
\Xi^{\star}\left(\tau,z\right)=\max_{\xi}\left(\bar{\Xi}\left(\tau,\xi,z\right)\right) .
\]
This is analogous to aligning the two sunquake signals along the horizontal distance axis. A best fit is identified based on which time-shift and excitation depth parameters maximize the cross-correlation and the ratio of maximum velocities within the first bounce wavepacket of the observed sunquake and model sunquake - which has the identified parameters - is used to obtain an estimate for the excitation mechanism's amplitude, which in turn is used to estimate transferred kinetic energy. 

We examine the results of the computation in the context of three main times: 1) When the time-derivative of SXR emission is at its maximum; 2) The time when the HXR emission is at its maximum; and 3) the start time of the sunquake as determined from strong perturbations in HMI so-called "bad pixels". In HMI Dopplergrams, there appear pixels which have wildly different values from their neighbors, and are often called bad pixels or anomalous pixels. These pixels appear as a failure of algorithms to correctly interpret filtergram data during Level 1 processing, and are associated with extreme values of Doppler velocity, magnetic field, and other observable which are associated with flare impacts in the low atmosphere \citep{HMI_doc}. We interpret these pixels as the excitation source location of sunquakes, and their first appearance as the start time (T$_{BP}$) of the sunquake.

The HXR and dSXR/dt peak times are obtained from KONUS-WIND data \citep{Aptekar1995,Lysenko2018}, which observes in three bands: G1 in the 21-82 keV range, G2 in the 82-331 keV range, and G3 in the 331-1252 keV range. The time derivative of the G1 band is used to find the dSXR/dt peak (T$_{G1}$) and the peak of the G3 curve is associated with the HXR peak (T$_{G3}$). These times, and the timing of bad pixel appearance, are listed in Table \ref{SQtable}; we note that bad pixels could not be identified in the M1.1 sunquake.

\begin{figure*}[htb]
\centering
\includegraphics[width=\linewidth]{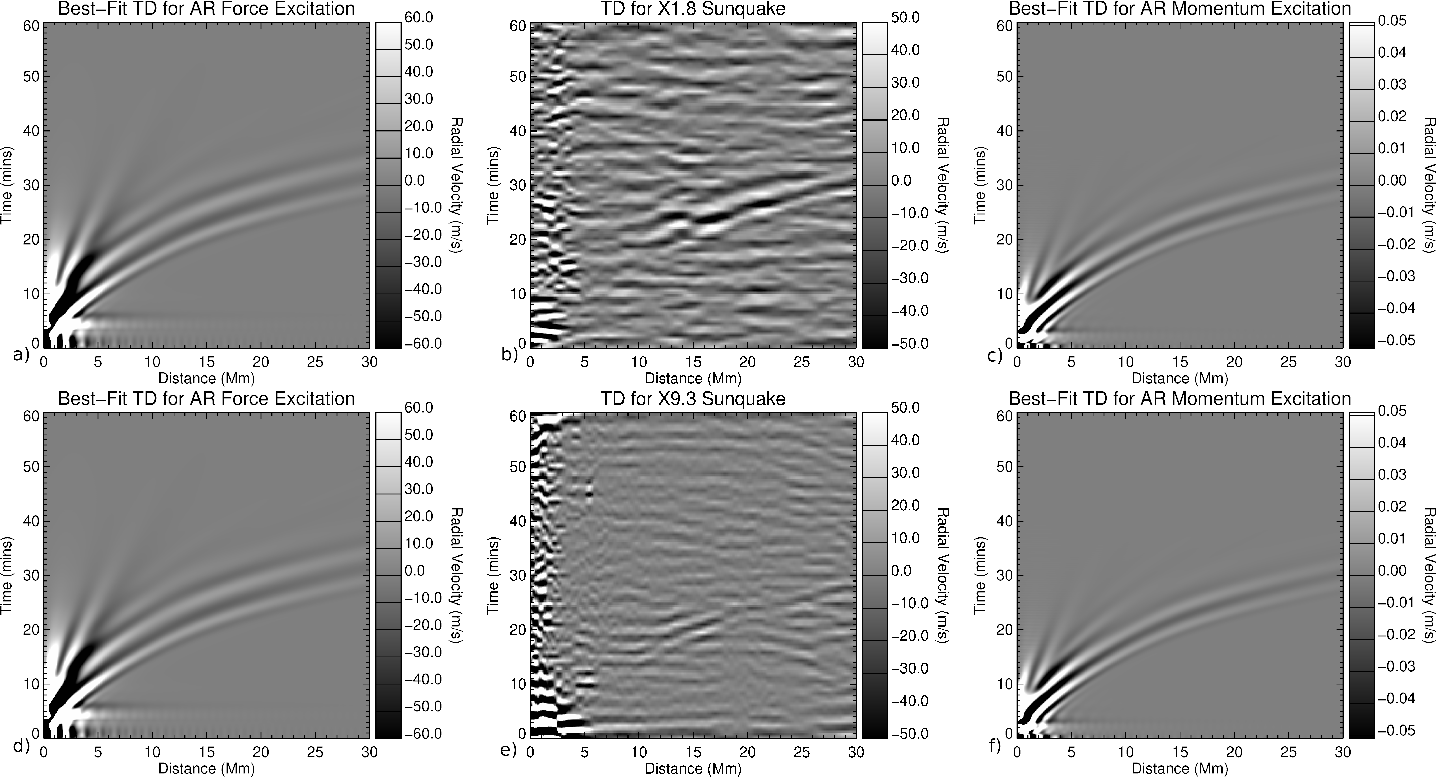}
\caption{Time-distance diagrams of: a) the best fit force case for the sunquake of the X1.8 flare; b) the sunquake produced by the X1.8 flare; c) the best fit momentum case for the sunquake of the X1.8 flare; d) the best fit force case for the sunquake of the X9.3 flare; e) the sunquake produced by the X9.3 flare; f) the best fit momentum case of the X9.3 flare. Darker pixels correspond to more negative velocities, lighter pixels correspond to more positive velocities.}
\label{comparison}
\end{figure*}

\begin{figure*}[!h]
\centering
\includegraphics[width=\linewidth]{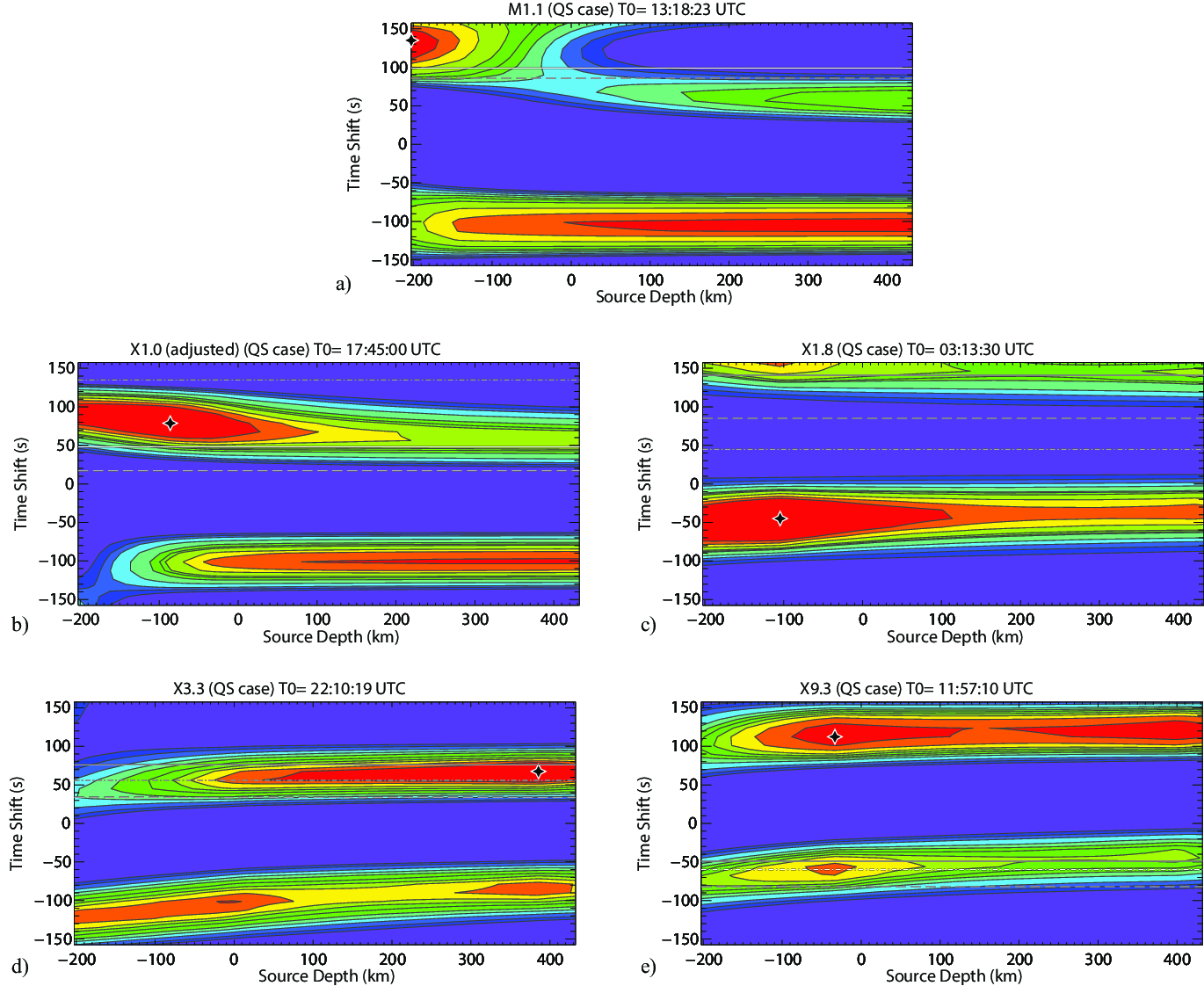}
\caption{Cross-correlation functions with the quiet sun damped momentum excitation model set. (a) Sunquake associated with the M1.1 flare; (b) Sunquake associated with the X1.0 flare; (c) Sunquake associated with the X1.8 flare; (d) Sunquake associated with the X3.3 flare; (e) Sunquake associated with the X9.3 flare. The contours begin at the median value, and each successive contour represents an increase in 5 percentile points (i.e. 50th, 55th, 60th, etc. percentiles). The solid horizontal line shows the HXR peak time, the dashed horizontal line shows the dSXR/dt peak time, and the dot-dashed horizontal line shows the suspected sunquake start time based on bad pixel count. The white and black diamond indicates where the parameters produce the greatest cross-correlation, and is used for energy estimation. Redder colors indicate greater correlation, green representing intermediate correlation, and purple representing low correlation.}
\label{vel_xcor}
\end{figure*}

\begin{figure*}[!h]
\centering
\includegraphics[width=\linewidth]{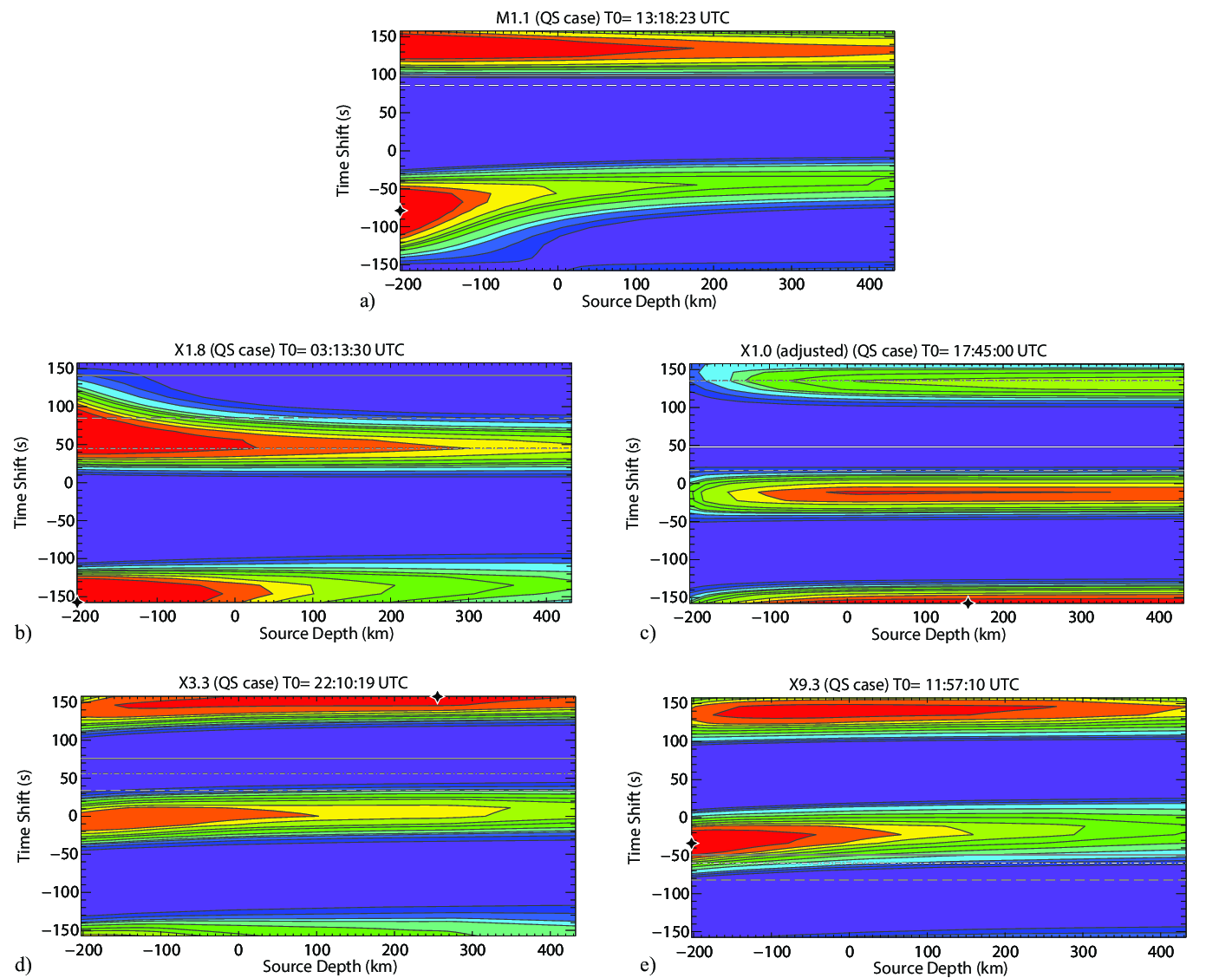}
\caption{The same as in Fig.~\ref{vel_xcor} for the force excitation.
}
\label{for_xcor}
\end{figure*}

The best fit parameters from the cross-correlation analysis using the momentum and force mechanisms are listed in Tables \ref{xcor_tableV} and \ref{xcor_tableF}, respectively; a visual comparison between the X1.8 sunquake and the identified force and momentum best fit models, and a comparison between the X9.3 sunquake and the identified best fits, are both shown in Figure \ref{comparison}. The two-dimensional cross-correlations for the quiet-sun damped momentum and force mechanisms are displayed in Figures \ref{vel_xcor} and \ref{for_xcor}, respectively. As a result of the periodicity in the signal, more than one band of best fit is often present. For momentum mechanisms, a majority of undamped sunquakes have energies bounded by $\num{1e28}$ ergs and excitation amplitudes on the order of 10 km s$^{-1}$. There does not seem to be an easily identifiable height at which these types of mechanisms may excite sunquakes, though the excitation's height consistently decreases when damping is increased.

When the observed sunquake events are compared with the force excitation set, the same relationship between excitation height and damping intensity remains. The amount of energy required to excite sunquakes in this way is also on the order $\num{1e28}$ ergs, and the energy estimates are more regular as damping is increased. Additionally, the bands of best fit for the force excitation set are offset relative to those of the momentum excitation set. This is consistent with the previous finding that force and momentum excitations produce sunquakes offset by two to three minutes, and is related to the -90 degree phase difference between the two signals.

The cross-correlation method does produce bands in close proximity to time shifts related to important moments in flare evolution, such as the HXR peak time. In two of the five sunquake events compared with the undamped momentum set, the best-fit case begins immediately following the HXR peak (the M1.1 and X1.0 event). This count increases to three in the quiet-sun case (including the X3.3 event), and reduces to 2 for active region damping. For force excitations, the undamped case has two best-fit cases close to the HXR peak, and one in the quiet sun and active region damping cases.

\section{Discussion and Conclusions}

It is expected that the estimated sunquake excitation location moves downwards as damping is increased, since excitement of acoustic waves in these lower layers will increase their amplitude once the waves reach the lower densities of higher atmospheric layers. This plays a role in the large amounts of energy required to excite the sunquakes in heavily-damped regions, as well as the increase in excitation amplitude needed to produce acoustic waves with the observed amplitudes. In particular, the active region damping is not necessarily representative of conditions for some sunquakes, as the observed waves travel outside the regions of strong magnetic field. The remaining sunquake events were excited at the edge of the respective active regions and propagated towards regions of lower magnetic field. The active region damping is particularly applicable in sunquake events where the acoustic waves travel across sunspot or active regions, though these types of events are difficult to observe and were not included in this study.

Additionally, the energy estimates are reliable only when the identified excitation coincides with or follows the HXR peak or bad pixel times, when the electron beam hypothesis is valid; we do not expect sunquake events to precede the HXR peak, which is a clear indication of energy deposition. The cases where this criteria is met include the M1.1, X1.0, and X3.3 events in the case of an instantaneous momentum excitation, and the M1.1 and X9.3 events in the case of an external force excitation. For both mechanisms, the X1.8 has a band of best fit in conjunction with the HXR, dSXR/dt, and bad pixel times, but no points in the band maximize the cross-correlation.

In the successfully analyzed events of momentum mechanisms, the energy required to excite the quake without damping is on order of $\num{1e28}$ ergs, consistent with recent estimates of \citet{Riuzhu} using acoustic holography methods. Furthermore, the energy estimate of a force mechanism for the M1.1 event is nearly equivalent to the momentum mechanism counterpart. In the case of the X3.3 event, the undamped momentum mechanism timing is not coincident with the HXR or bad pixel times, though increasing the damping temporally aligns the excitation nearly exactly with the HXR peak time. This quiet sun damping case also gives an energy estimate on the order of $\num{1e28}$ ergs, up to a maximum of $\num{1e29}$ ergs for the active region damped case which also provides an excitation coincident with the HXR peak time. Similar circumstances arise for the X9.3 case with an external force mechanism, where the undamped case is unrelated temporally with any main times but the quiet sun damping indicates an excitation time just following the HXR peak time. The energy estimate for the X9.3 force excitation with quiet sun damping is roughly $\num{1e29}$ ergs; this is not unreasonable for such a strong flare, which may release up to $\num{1e32}$ ergs \citep{Xenergy}.

Moving forward, it is clear that sunquake signals are degenerate in parameter space with respect to source depth, and also time shift. The time shift degeneracy is relatively easy to deal with, as we expect the moment of excitation to have timing close to the X-ray and bad pixel times, and this can be accounted for. The excitation depth degeneracy is more difficult to treat, as lower-amplitude excitations produced by deep sources can be compensated for by greater energy deposition. There are also several events - notably the X1.0 event - which indicate the presence of a high excitation source, though such excitations tend to produce weaker p-modes and stronger acoustic-gravity waves in the model as the height increases. Further study of atmospheric acoustic-gravity waves is necessary to understand their role in sunquake excitation and propagation.

In conclusion, we find that at least three sunquake events for the momentum mechanism and at least two events for the force mechanism are consistent with the electron beam hypothesis. The excitation start times of these events are coincident with or closely follow the time of peak HXR emission, which is a reliable diagnostic of energy deposition. In these cases, the energy required to excite the sunquakes falls within expectations based on previous studies, and in some cases indicates a moderate amount of acoustic damping in the region of sunquake propagation.

\acknowledgements
The research was supported by NASA grants NNX14AB68G, NNX16AP05H, and NSF grant 1916509. 

\appendix
\section{Stability Analysis of the Numerical Finite-Difference Method}\label{stab}

Begin with discretizing the governing equations and substituting associated errors, keeping in mind that we use a staggered grid, where the half-grid variables are linearly interpolated onto a given grid point, i.e.

\[
v_{r,i}^{n}=v_{r,i-\frac{1}{2}}^{n}+\dfrac{v_{r,i+\frac{1}{2}}^{n}-v_{r,i-\frac{1}{2}}^{n}}{\Delta r_{i}}\dfrac{\Delta r_{i}}{2}=v_{r,i-\frac{1}{2}}^{n}+\dfrac{v_{r,i+\frac{1}{2}}^{n}-v_{r,i-\frac{1}{2}}^{n}}{2}
\]

\[\left\lbrace\begin{array}{l}
\dfrac{\partial v_{h}}{\partial t} = -\dfrac{P^{\prime}}{r \rho_{0}} \\

\dfrac{\partial \bar{\rho}}{\partial t}+v_{r}\dfrac{\partial \ln\rho_{0}}{\partial r}+\dfrac{2v_{r}}{r}+\dfrac{\partial v_{r}}{\partial r}-\dfrac{L^{2}}{r}v_{h}=0 \\

\dfrac{\partial v_{r}}{\partial t}=-\dfrac{1}{\rho_{0}}\dfrac{\partial P^{\prime}}{\partial r}+\bar{\rho}g \\

\dfrac{\partial P^{\prime}}{\partial t} = -\gamma P_{0}\left(\left(\dfrac{N^{2}}{g}+\dfrac{\partial \ln \rho_{0}}{\partial r}+\dfrac{2}{r}\right)v_{r}+\dfrac{\partial v_{r}}{\partial r}-\dfrac{L^{2}}{r}v_{h}\right)
\end{array}\right.
\]
\[
\downarrow
\]
\[
\left\lbrace\begin{array}{l}
v_{h,i}^{n+1}=v^{n}_{h,i}-\dfrac{\Delta t}{r\rho_{0}}P_{i}^{n+1}
\\
\dfrac{\rho_{i}^{n+1}-\rho_{i}^{n}}{\Delta t}=-\dfrac{1}{2}x_{1}\left(v_{r,i+\frac{1}{2}}^{n}+v_{r,i-\frac{1}{2}}^{n}\right)-\dfrac{1}{12\Delta r}\left(v_{r,i-\frac{3}{2}}^{n}-8v_{r,i-\frac{1}{2}}^{n}+8v_{r,i+\frac{1}{2}}^{n}-v_{r,i+\frac{3}{2}}^{n}\right)+\dfrac{L^{2}}{r}v_{h,i}^{n}
\\
\dfrac{v_{r,i}^{n+1}-v_{r,i}^{n}}{\Delta t} = -\dfrac{1}{12\rho_{0}\Delta r}\left(P_{i-\frac{3}{2}}^{n+1}-8P_{i-\frac{1}{2}}^{n+1}+8P_{i+\frac{1}{2}}^{n+1}-P_{i+\frac{3}{2}}^{n+1}\right)+\dfrac{1}{2}\left(\bar{\rho}^{n+1}_{i-\frac{1}{2}}+\bar{\rho}^{n+1}_{i+\frac{1}{2}}\right)g
\\
\dfrac{P_{i}^{n+1}-P_{i}^{n}}{\Delta t} = \gamma P_{0}\left[-\dfrac{1}{2}\left(v_{r,i+\frac{1}{2}}^{n}+v_{r,i-\frac{1}{2}}^{n}\right)x_{2}-\dfrac{1}{12\Delta r}\left(v_{r,i-\frac{3}{2}}^{n}-8v_{r,i-\frac{1}{2}}^{n}+8v_{r,i+\frac{1}{2}}^{n}-v_{r,i+\frac{3}{2}}^{n}\right)+\dfrac{L^{2}}{r}v_{h,i}^{n}\right]\end{array}\right.
\]

where $x_{1}=\partial \ln \rho_{0}/\partial r+2/r$ and $x_{2}=x_{1}+N^{2}/g$. The scheme is semi-implicit, as the updated values (of the $(n+1)^{th}$ time-step) are used to determine the new values of radial and horizontal velocity. The error associated with $v_{h}$ can be expressed in terms of Fourier components
\[
v_{h,i}^{n}=\sum_{m}D_{m}e^{a_{m}t_{n}}e^{ik_{m}r_{i}}
\]
The errors associated with $\bar{\rho}$, $P^{\prime}$, and $v_{r}$ have the same form with coefficients $A$, $B$, and $C$, respectively. Let us consider a single Fourier mode ($m=m^{\prime}$) and divide through by $e^{ik r_{i}}e^{at_{n}}$
\[
\left\lbrace\begin{array}{l}
GD=D-G\dfrac{\Delta t B}{r\rho_{0}}
\\
GA=A+\Delta t\left[-\dfrac{C}{2}x_{1}\left(e^{\frac{ik\Delta r}{2}}+e^{-\frac{ik\Delta r}{2}}\right)-\dfrac{C}{12\Delta r}\left(e^{-\frac{3ik\Delta r}{2}}-8e^{-\frac{ik\Delta r}{2}}+8e^{\frac{ik\Delta r}{2}}-e^{\frac{3ik\Delta r}{2}}\right)+\dfrac{DL^{2}}{r}\right]
\\
GC = C +G\Delta t\left[ -\dfrac{B}{12\rho_{0}\Delta r}\left(e^{-\frac{3ik\Delta r}{2}}-8e^{-\frac{ik\Delta r}{2}}+8e^{\frac{ik\Delta r}{2}}-e^{\frac{3ik\Delta r}{2}}\right)+\dfrac{A}{2}\left(e^{\frac{ik\Delta r}{2}}+e^{-\frac{ik\Delta r}{2}}\right)g\right]
\\
GB=B+\Delta t \gamma P_{0}\left[-\dfrac{C}{2}\left(e^{\frac{ik\Delta r}{2}}+e^{-\frac{ik\Delta r}{2}}\right)x_{2}-\dfrac{C}{12\Delta r}\left(e^{-\frac{3ik\Delta r}{2}}-8e^{-\frac{ik\Delta r}{2}}+8e^{\frac{ik\Delta r}{2}}-e^{\frac{3ik\Delta r}{2}}\right)+\dfrac{DL^{2}}{r}\right]
\end{array}\right.
\]
For convenience in evaluating the characteristic equation, we define
\[
x_{3}=\dfrac{1}{2}\left(e^{\frac{ik\Delta r}{2}}+e^{-\frac{ik\Delta r}{2}}\right)=\cos\left(\frac{k\Delta r}{2}\right)
\]
\[
x_{4}=\dfrac{1}{12}\left(e^{-\frac{3ik\Delta r}{2}}-8e^{-\frac{ik\Delta r}{2}}+8e^{\frac{ik\Delta r}{2}}-e^{\frac{3ik\Delta r}{2}}\right)=\dfrac{i}{6}\left(8\sin\left(\frac{k\Delta r}{2}\right)-\sin\left(\frac{3k\Delta r}{2}\right)\right)
\]

We evaluate the stability at the shortest wavelength ($\lambda=2\Delta r$, $k=\pi/\Delta r$), which is usually the most unstable oscillation mode. Substituting the value for $k$ in $x_{3}$ and $x_{4}$ yields $x_{3}=0$ and $x_{4}=3i/2$.

In order to express the stability of the system in terms of non-dimensional parameters, we define new Fourier coefficients $A^{\prime}, B^{\prime}, C^{\prime}$ and distribute the appropriate variables

\[
A^{\prime}=Ac \leftrightarrow A=\dfrac{A^{\prime}}{c}
\]
\[
B^{\prime}= \dfrac{B}{\rho_{0}c} \leftrightarrow B=\rho_{0}cB^{\prime}
\]
\[
C^{\prime} = C
\]
\[
D^{\prime}=D
\]
Additionally, the dimensionless parameters we wish to express the system in are

\[
N_{c}=\dfrac{c\Delta t}{\Delta r}
\]
\[
N_{s}=\dfrac{c\Delta t}{r}
\]
\[
N_{g}=\dfrac{g\Delta t}{c}
\]
\[
N_{\rho}=\dfrac{c\Delta t}{H_{\rho}}=c\Delta t \dfrac{\partial \ln \rho_{0}}{\partial r}
\]
\[
N_{N}=\dfrac{c\Delta t N^{2}}{g}
\]

Where $N_{c}$ is the Courant number, the ratio between the speed of sound waves and the speed of the fastest radial wave allowed by the system. $N_{s}$ is the ratio of the distance traveled by a radially propagating pressure wave to its given radial position. When $N_s$ is multiplied by $L^{2}$, it is the same ratio but now in terms of horizontally propagating waves. $N_{g}$ is the ratio between the sound speed and the change in velocity experienced by a fluid element due to gravity per unit time. $N_{\rho}$ is the ratio of the distance traveled by a radially propagating wave to the density scale height, and $N_{N}$ is a stability parameter for gravity waves.

Beginning with the equation for perturbed density and multiplying everywhere by $c$

\[
c\left[GA=A-\Delta t x_{1}x_{3}C-\dfrac{\Delta t}{\Delta r}Cx_{4}+\dfrac{\Delta t L^{2}}{r}D\right]
\]
\[
GA^{\prime}=A^{\prime}-\left(2\dfrac{c\Delta t}{r}-\dfrac{c \Delta t}{H_{\rho}}\right)Cx_{3}-\dfrac{c\Delta t}{\Delta r}Cx_{4}+\dfrac{c\Delta t}{r}L^{2}D
\]
\[
\boxed{GA^{\prime}=A^{\prime}-\left(2N_{s}-N_{\rho}\right)Cx_{3}-N_{c}Cx_{4}+N_{s}L^{2}D}
\]
Moving to the equation for perturbed pressure and multiplying everywhere by $1/(\rho_{0}c)$

\[
\dfrac{1}{\rho_{0}c}\left[GB=B-\gamma P_{0}\Delta t x_{2}x_{3}C-\gamma P_{0}\dfrac{\Delta t}{\Delta r}x_{4}C+\gamma P_{0}\dfrac{\Delta t}{r}L^{2}D\right] 
\]

\[
GB^{\prime}=B^{\prime}-c\Delta t\left(\dfrac{2}{r}-\dfrac{1}{H_{\rho}}+\dfrac{N^{2}}{g}\right)x_{3}C-\dfrac{c\Delta t}{\Delta r}x_{4}C+\dfrac{c\Delta t}{r}L^{2}D
\]

\[
\boxed{GB^{\prime}=B^{\prime}-\left(2N_{s}-N_{\rho}+N_{N}\right)x_{3}C-N_{c}x_{4}C+N_{s}L^{2}D}
\]
Substituting the new Fourier coefficients $A^{\prime}$ and $B^{\prime}$ into the radial momentum equation yields

\[
GC=C-\dfrac{G}{\rho_{0}}\dfrac{\Delta t}{\Delta r}x_{4}B-Gx_{3}g\Delta t A
\]

\[
GC=C-G\dfrac{c\Delta t}{\Delta r}x_{4}B^{\prime}-G\dfrac{g\Delta t}{c}x_{3}A^{\prime}
\]

\[
\boxed{G\left(C+N_{c}x_{4}B^{\prime}+N_{g}x_{3}A^{\prime}\right)=C}
\]
And performing a similar substitution in the horizontal momentum equation yields

\[
GD=D-G\dfrac{\Delta t}{r}\dfrac{1}{\rho_{0}}B
\]

\[
GD=D-G\dfrac{\Delta t}{r}\dfrac{1}{c}B^{\prime}
\]

\[
\boxed{G\left(D+N_{s}B^{\prime}\right)=D}
\]

In general, the above conditions form the matrix equation $GY\mathbf{x}=Z\mathbf{x}$, or

\[
G
\begin{bmatrix}
1 & 0 & 0 & 0 \\
0 & 1 & 0 & 0 \\
x_{3}N_{g} & x_{4}N_{c} & 1 & 0 \\
0 & N_{s} & 0 & 1
\end{bmatrix}
\begin{bmatrix}
A^{\prime} \\
B^{\prime} \\
C \\
D
\end{bmatrix}
=
\begin{bmatrix}
1 & 0 & -\left(2N_{s}-N_{\rho}\right)x_{3}-x_{4}N_{c} & L^{2}N_{s} \\
0 & 1 & -\left(2N_{s}-N_{\rho}+N_{N}\right)x_{3}-x_{4}N_{c} & L^{2}N_{s} \\
0 & 0 & 1 & 0 \\
0 & 0 & 0 & 1
\end{bmatrix}
\begin{bmatrix}
A^{\prime} \\
B^{\prime} \\
C \\
D
\end{bmatrix}
\]

\begin{figure}[htb]
\begin{center}
\includegraphics[width=\linewidth]{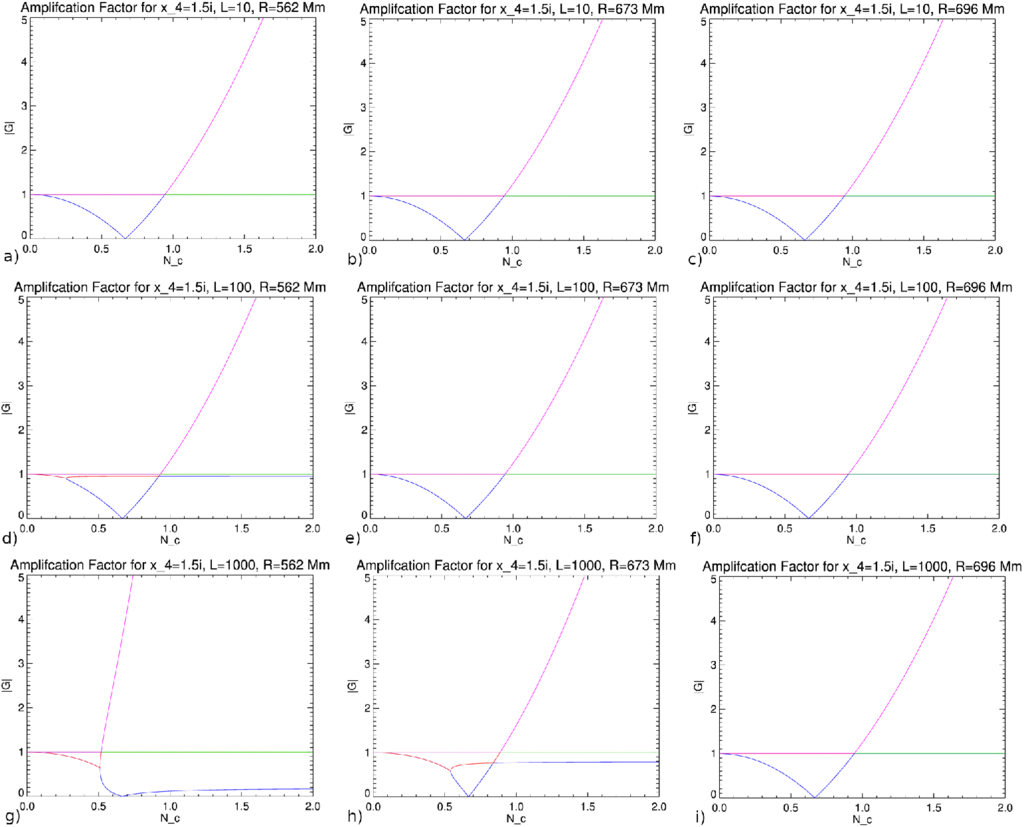}
\caption{Amplitude of amplification factors for $l=0$ at $R=562$ Mm (a), at $R=673$ Mm (b), and at $R=696$ Mm (c); for $l= 100$ at $R=562$ Mm (d), at $R=673$ Mm (e), and at $R=696$ Mm (f); for $l=1000$ at $R=562$ Mm (g), at $R=673$ Mm (h), and at $R=696$ Mm (i).}\label{stab_pic}
\end{center}
\end{figure}

Solving the above system for $G$ yields the amplification factors, whose amplitudes are bounded by 1 for all values of $l$. Figures \ref{stab_pic} (a,b,c), (d,e,f), and (g,h,i) show how the amplitude of the amplification factors grow with the Courant number $N_{c}$, for $l=0$, $l=100$, and $l=1000$ respectively. The parameters in the above system are evaluated locally, for $R=562$ Mm (Figures \ref{stab_pic} a,d,g), $R=673$ Mm (Figures \ref{stab_pic} b,e,h), and $R=696$ Mm (Figures \ref{stab_pic} c,f,i). While the $l=1000$ mode is unstable at $R=562$ Mm for our Courant number ($N_{c}=0.6$), this instability is more physical than numerical. This mode has a lower turning point of $R=664$ Mm, so $R=562$ Mm is never reached; in other words, we should require the system to be stable only in a given modes region of propagation, which is indeed the case.

\newpage

\bibliography{mybib}

\newpage

\newpage
\newpage

\newpage

\newpage
\newpage

\newpage
\newpage

\begin{table*}[!ht]
\begin{center}
\caption{Best Fit Parameters for Sunquake Events - Momentum Mechanism}\label{xcor_tableV}
\begin{tabular}{||c|c||c|c|c|c|c||}
\hline
\multicolumn{2}{||c||}{} &
\multicolumn{5}{c||}{Momentum Case}
\\
\hline
\multicolumn{1}{||c|}{Flare} &
\multicolumn{1}{c||}{Damping} &
T$_{shift}$ (s) & Height (km)  & Amp. (g cm s$^{-1}$) & Max. V (km s$^{-1}$) & Energy (ergs) 
\\
\hline\hline
X1.8 & Undamped & -45 & -17 & $\num{1.95e22}$ &11.1 & $\num{6.39e27}$ \\
\hline
& Quiet Sun & -45 & -104 & $\num{3.31e23}$ & 141.8 & $\num{1.39e30}$  \\
\hline
& Active Region & -56.25 & -142 & $\num{1.76e24}$ & 675.6 & $\num{3.53e31}$  \\
\hline
X9.3 & Undamped & +112.5 & +432 & $\num{5.40e21}$ & 32.5 & $\num{3.05e27}$  \\
\hline
& Quiet Sun & +112.5 & -33 & $\num{1.54e23}$ & 82.3 & $\num{3.74e29}$ \\
\hline
& Active Region & +112.5 & -68 & $\num{1.54e24}$ & 785.8 & $\num{3.82e31}$ \\
\hline
X3.3 & Undamped & +135 & +386 & $\num{2.68e21}$ & 11.5 & $\num{5.88e26}$ \\
\hline
& Quiet Sun & +67.5 & +386 & $\num{2.83e22}$ & 121.2 & $\num{6.57e28}$  \\
\hline
& Active Region & -123.75 & -162 & $\num{1.65e24}$ & 599.8 & $\num{2.94e31}$\\
\hline
X1.0 & Undamped & +67.5 & +87 & $\num{2.87e22}$ & 24.0 & $\num{1.94e28}$  \\
\hline
& Quiet Sun & +78.75 & -86 & $\num{9.48e23}$ & 430.2 & $\num{1.21e31}$  \\
\hline
& Active Region & +78.75 & -123 & $\num{6.96e24}$ & 2825.6 & $\num{5.85e32}$  \\
\hline
M1.1 & Undamped & +101.25 & -182 & $\num{4.60e22}$ & 15.8 & $\num{2.15e28}$  \\
\hline
& Quiet Sun & +135 & -203 & $\num{7.34e23}$ & 238.8 & $\num{5.19e30}$  \\
\hline
& Active Region & +123.75 & -203 & $\num{5.47e24}$ & 1780.1 & $\num{2.88e32}$ \\
\hline\hline
\end{tabular}
\end{center}
\end{table*}
\newpage
\begin{table*}[htb]
\begin{center}
\caption{Best Fit Parameters for Sunquake Events - Force Mechanism}\label{xcor_tableF}
\begin{tabular}{||c|c||c|c|c|c||}
\hline
\multicolumn{2}{||c||}{} &
\multicolumn{4}{c||}{Force Case}
\\
\hline
\multicolumn{1}{||c|}{Flare} &
\multicolumn{1}{c||}{Damping} &
T$_{shift}$ (s) & Height (km) & Amp. (dyn cm$^{-3}$) & Energy (ergs)
\\
\hline\hline
X1.8 & Undamped & -146.25 & -203 & $\num{1.38e-2}$ & $\num{1.01e28}$ \\
\hline
& Quiet Sun & -157.5 & -203 & $\num{1.01e-1}$ & $\num{5.58e28}$ \\
\hline
& Active Region & -157.5 & -203 & $\num{6.59e-1}$ & $\num{2.58e29}$ \\
\hline
X9.3 & Undamped & +135 & +129 & $\num{2.19e-3}$ & $\num{2.27e27}$ \\
\hline
& Quiet Sun  & -33.75 & -203 & $\num{2.29e-1}$ & $\num{1.27e29}$ \\
\hline
& Active Region  & -45 & -203 & 1.57 & $\num{6.17e29}$ \\
\hline
X3.3 & Undamped & +135 & +181 & $\num{5.65e-4}$ & $\num{6.44e26}$ \\
\hline
& Quiet Sun & +157.5 & +255 & $\num{1.30e-2}$ & $\num{1.32e28}$ \\
\hline
& Active Region &0 & -33 & $\num{4.24e-1}$ & $\num{2.07e29}$ \\
\hline
X1.0 & Undamped & -157.5 & +327 & $\num{3.59e-3}$ & $\num{4.56e27}$ \\
\hline
& Quiet Sun  & -157.5 & +155 & $\num{5.34e-2}$ & $\num{4.73e28}$ \\
\hline
& Active Region & -11.25 & +432 & $\num{6.52e-1}$ & $\num{3.73e29}$ \\
\hline
M1.1 & Undamped & +146.25 & -203 & $\num{4.04e-2}$ & $\num{2.93e28}$ \\
\hline
& Quiet Sun & -78.5 & -203 & $\num{2.59e-1}$ & $\num{1.44e29}$ \\
\hline
& Active Region & -90 & -203 & 1.45 & $\num{5.68e29}$\\
\hline
\end{tabular}
\end{center}
\end{table*}

\end{document}